\documentclass[aps,prl,nofootinbib,twocolumn,superscriptaddress,floatfix,notitlepage]{revtex4-1}
\pdfoutput=1

\usepackage{graphicx}
\usepackage{amsmath,amsfonts,amssymb}
\usepackage[dvipsnames]{xcolor}
\usepackage[breaklinks,colorlinks,urlcolor=blue,citecolor=blue,linkcolor=magenta]{hyperref}
\usepackage{verbatim}
\usepackage{enumitem}
\usepackage{hyperref}
\usepackage{soul}
\hypersetup{urlcolor=PineGreen,
	        citecolor=PineGreen,
	        linkcolor=PineGreen}

\usepackage{aas_macros}

\newcommand{\td}{{\rm d}}

\newcommand{\msun}{{M_\odot}}

\newcommand{\vect}[1]{\boldsymbol{#1}}

\newcommand{\be}{\begin{equation}}
\newcommand{\ee}{\end{equation}}
\newcommand{\bea}{\begin{equation} \begin{aligned}}
\newcommand{\eea}{\end{aligned} \end{equation}}

\def\lsim{\mathrel{\raise.3ex\hbox{$<$\kern-.75em\lower1ex\hbox{$\sim$}}}}
\def\gsim{\mathrel{\raise.3ex\hbox{$>$\kern-.75em\lower1ex\hbox{$\sim$}}}}

\newcommand{\papertitle}{Fuzzy dark matter fails to explain the dark matter cores}

\begin{document}

\title{\papertitle}

\author{María Benito}
\affiliation{Tartu Observatory, University of Tartu, Observatooriumi 1, Toravere 61602, Estonia}

\author{Gert H\"utsi}
\affiliation{Keemilise ja Bioloogilise F\"u\"usika Instituut, R\"avala pst. 10, 10143 Tallinn, Estonia}

\author{Kristjan Müürsepp}
\affiliation{Keemilise ja Bioloogilise F\"u\"usika Instituut, R\"avala pst. 10, 10143 Tallinn, Estonia}
\affiliation{ INFN, Laboratori Nazionali di Frascati, C.P.~13, 100044 Frascati, Italy}

\author{Jorge Sánchez~Almeida}
\affiliation{Instituto de Astrof\'\i sica de Canarias, La Laguna, Tenerife, E-38200, Spain}
\affiliation{Departamento de Astrof\'\i sica, Universidad de La Laguna, Spain}

\author{Juan Urrutia}
\email{juan.urrutia@kbfi.ee}
\affiliation{Keemilise ja Bioloogilise F\"u\"usika Instituut, R\"avala pst. 10, 10143 Tallinn, Estonia}
\affiliation{Departament of Cybernetics, Tallinn University of Technology, Akadeemia tee 21, 12618 Tallinn, Estonia}

\author{Ville Vaskonen}
\affiliation{Keemilise ja Bioloogilise F\"u\"usika Instituut, R\"avala pst. 10, 10143 Tallinn, Estonia}
\affiliation{Dipartimento di Fisica e Astronomia, Universit\`a degli Studi di Padova, Via Marzolo 8, 35131 Padova, Italy}
\affiliation{INFN, Sezione di Padova, Via Marzolo 8, 35131 Padova, Italy}

\author{Hardi Veerm\"ae}
\affiliation{Keemilise ja Bioloogilise F\"u\"usika Instituut, R\"avala pst. 10, 10143 Tallinn, Estonia}

\begin{abstract}
Ultrafaint dwarf galaxies (UFDs) are ideal for studying dark matter (DM) due to minimal baryonic effects. UFD observations suggest cored DM profiles. We find that the core radius -- stellar mass scaling predicted by fuzzy dark matter (FDM) is at $6.1\sigma$ tension with UFD observations. Combining observations from 27 UFDs, the required FDM mass $m_a = 3.2_{-0.6}^{+0.8}\times 10^{-21}\,{\rm eV}$ is also in conflict with existing Lyman-$\alpha$ bounds. Our results suggest that FDM cannot provide a consistent explanation for DM cores and imply $m_a > 2.2\times 10^{-21}\,{\rm eV}$ at to $2\sigma$ CL.
\end{abstract}

\maketitle

\vspace{5pt}\noindent\textbf{Introduction -- } The nature of dark matter (DM) is one of the biggest open questions in cosmology. It can range from ultralight bosons of mass $\gtrsim 10^{-22}\,{\rm eV}$~\cite{Hu:2000ke,Marsh:2015xka,Hui:2016ltb} to macroscopic $\mathcal{O}(1\,\msun)$ objects~\cite{Carr:2020xqk}. In the standard cold dark matter (CDM) paradigm, DM consists of collisionless non-relativistic particles with negligible non-gravitational interactions~\cite{Planck:2018vyg, Petac:2019lam, Chabanier:2019eai}. This simple but successful hypothesis faces challenges on small scales, such as the cusp/core, missing satellites, and too-big-to-fail problems (see e.g.~\cite{Bullock:2017xww, Perivolaropoulos:2021jda}). Although many of these discrepancies may be addressed by considering the role of baryonic feedback, fuzzy dark matter (FDM) models offer an effective alternative solution also in small halos where baryonic effects can be insignificant. In these models, DM consists of ultralight bosons that behave as classical waves on galactic scales. Consequently, FDM halos exhibit a flat core~\cite{Chan:2021bja} in contrast to pure CDM that predicts cuspy central regions~\cite{Moore:1994yx,2015AJ....149..180O,Laporte:2013fwa,Oh:2008ww}. Additionally, the wave nature of FDM suppresses the abundance of light halos, reducing the number of satellite galaxies~\cite{Moore:1999nt, Klypin:1999uc,2012MNRAS.422.1203B}. 

In galaxies with stellar masses $M_* \gtrsim 10^6\,\msun$, baryonic feedback effects may be responsible for the observed cored DM profiles~\cite{Mashchenko:2007jp, Pontzen:2011ty, Teyssier:2012ie, Read:2015sta, Tollet:2015gqa}. In less massive galaxies, the energy released by stars is insufficient to modify the distribution of DM and stellar feedback is not strong enough to produce cored DM profiles \cite{2020MNRAS.497.2393L,2016MNRAS.456.3542T}. Additionally, these halos are so light that photo-heating from the reionizing UV background severely suppresses their star formation efficiency~\cite{Somerville:2001km,2015ARA&A..53...51S}. Thus, for these $M_\star \lesssim 10^6\,\msun$ galaxies, the shape of the DM halo is believed to be set by the DM properties. 

In this letter, we analyse FDM by accounting for observations of ultralight dwarf galaxies (UFDs) in which baryonic feedback effects are insignificant~\cite{Lazar:2020pjs, Onorbe:2015ija, 2012ApJ...759L..42P, 2020ApJ...904...45H, 2021MNRAS.502.4262J} and have been used to test FDM~\cite{Marsh:2018zyw, Hayashi:2021xxu, Dalal:2022rmp, Zimmermann:2024xvd, Teodori:2025rul}. We find that a FDM explanation of the stellar cores in UFDs implies a DM mass $3.2_{-0.6}^{+0.8}\times10^{-21}{\rm eV}$, which is excluded for example by the Lyman-$\alpha$ forest observations~\cite{Rogers:2020ltq}. Using the lightest UFDs, we place a new bound on the DM mass $m_a > 2.2\times 10^{-21}\,{\rm eV}$ at $2\sigma$ CL. In agreement with~\cite {Burkert:2020laq, Salucci:2020cmt}, we find that FDM predicts a negative core radius-stellar mass relationship, opposite to what is observed. Quantifying this by comparing the FDM prediction with a general power-law fit, we find that the significance of the tension is $6.1\, \sigma$. Overall, we aim to demonstrate that existing observations essentially exclude FDM as an explanation for the cored DM profiles.

\vspace{5pt}\noindent\textbf{Fuzzy Dark Matter halos -- } We analyze 27 UFDs whose observational data is compiled in~\cite{2008ApJ...684.1075M,DES:2015zwj,2018ApJ...860...66M,Almeida:2024cqa}. These galaxies, which are among the lightest satellites of the Milky Way and the Magellanic Clouds, have stellar masses of $\mathcal{O} (10^3-10^5)\,\msun$ and exhibit cored stellar profiles. The stellar core radii $r^{*}_c$ and the stellar masses $M_*$ of these UFDs are displayed in Table~\ref{tab:table_1}. The stellar masses of the first six UFDs are derived from photometry assuming a mass-to-light ratio of $2\,\msun/L_\odot$, as approximately expected given the observed stellar populations~\cite{2024ApJ...967...72R}. The quoted uncertainties contain contributions from both the photometry and the distance measurements. For the next 13 UFDs, we take the stellar masses from~\cite{2008ApJ...684.1075M} assuming a Salpeter initial stellar mass function and for the last 8, we use stellar masses from~\cite{DES:2015zwj}, which assumes a Chabrier initial stellar mass function.

Reconstructing the DM profiles of UFDs is challenging due to the lack of reliable dynamical measurements. Nonetheless, the application of the Eddington inversion method to the observed stellar distribution suggests that cored DM profiles are preferred over cuspy NFW profiles in the first 6 UFDs listed in Table~\ref{tab:table_1}~\cite{Almeida:2024cqa}.  We assume that this holds also for the other 21 UFDs. The evidence for cored DM profiles suggests deviations from the CDM paradigm, as it is believed that the DM cores cannot be attributed to baryonic feedback effects expected to be negligible in these systems.

\begin{table}
    \centering
        \begin{tabular}{l|c|c|c}
               & $M_\star/10^3\msun$ & $r_c^\star/\,{\rm pc}$ & $m_a/{10^{-21}\rm eV}$ \\
            \hline
            Horologium-I & $1.96\pm0.40$ & $24.3\pm5.0$&$8.7^{+20.5}_{-6.1}$ \\
            Horologium-II &$2.47\pm0.50$ &$22.5\pm4.6$&$8.8^{+20.6}_{-6.2}$\\
            Hydra-II &$7.1\pm1.46$& $61.0\pm12.5$& $3.0^{+6.2}_{-2.0}$\\
            Phoenix-II &$1.31\pm0.27$&$27.9\pm15.73$& $8.4^{+24.3}_{-6.3}$\\
            Sagittarius-II &$2.47\pm0.50$&$23.4\pm4.81$&$8.5^{+19.8}_{-5.9}$ \\
            Triangulum-II &$0.89\pm0.18$&$20.4\pm4.19$& $12.0^{+29.9}_{-8.6}$\\
            \hline
            Boötes-I  &$67\pm6$&$242^{+22}_{-20}$& $0.9^{+1.5}_{-0.6}$\\
            Boötes-II &$2.8^{+1.3}_{-1.0}$&$51\pm17$&$6.9^{+16.5}_{-4.8}$\\
            Canes Venatici-I  &$580\pm40$&$564\pm36$&$0.3^{+0.4}_{-0.2}$\\
            Canes Venatici-II &$16^{+4.0}_{-3.0}$&$74^{+14}_{-10}$&$3.4^{+6.8}_{-2.3}$\\
            Coma Bernices  &$9.2\pm1.7$&$77\pm10$&$3.7^{+7.7}_{-2.5}$\\
            Hercules  &$72^{+12}_{-11}$&$330^{+75}_{-52}$&$0.7^{+1.1}_{-0.4}$\\
            Leo-IV  &$16^{+6}_{-4}$&$116^{+26}_{-34}$&$2.3^{+4.7}_{-1.6}$\\
            Segue-I &$1.3\pm0.2$&$29^{+8}_{-5}$&$13.2^{33.0}_{-9.5}$\\
            Ursa Major-I  &$37^{+6}_{-5}$&$318^{+50}_{-39}$&$0.8^{+1.4}_{-0.5}$\\
            Ursa Major-II  &$12\pm 1$&$140\pm 25$&$2.1^{+4.1}_{-1.4}$\\
            Willman-I  &$3.2\pm 0.6$&$25^{+5}_{-6}$&$12.2^{+29.3}_{-8.6}$\\
            SDSSJ1058+2843  &$0.4^{+0.145}_{+0.120}$&$22.0^{+5.0}_{-4.0}$&$21.9^{+58.5}_{-15.9}$\\
            Draco  &$620\pm 10$&$221\pm 16$&$0.6^{+0.8}_{-0.3}$\\
            \hline
            Gru-II &$3.4^{+0.3}_{-0.4}$&$93\pm14$&$4.0^{+8.6}_{-2.7}$\\
            Tucana-III &$0.8\pm0.1$&$44\pm6$&$10.5^{+25.5}_{-7.4}$\\
            Tucana-IV &$2.2^{+0.4}_{-0.3}$&$127\pm24$&$3.4^{+7.3}_{-2.3}$ \\
            Columba-I &$6.2^{+1.9}_{-1.0}$&$103\pm25$& $3.2^{+6.7}_{-2.1}$\\
            Reticilum-III &$2.0^{+0.6}_{-0.7}$&$64\pm24$& $6.1^{+15.0}_{-4.4}$\\
            Tucana-V &$0.5\pm0.1$&$17\pm6$& $25.7^{+72.5}_{-18.9}$\\
            Indus-II &$4.9^{+1.8}_{-1.6}$&$181\pm67$&$2.1^{+4.5}_{-1.4}$ \\
            Cetus-II &$0.1\pm{+0.04}$&$17\pm7$& $37.4^{+118.9}_{-28.5}$\\
            \hline
            combined & & & $3.2_{-0.6}^{+0.8}$ 
        \end{tabular}
    \caption{Observed stellar masses and stellar core radii of the 27 UFDs considered in this study. The stellar radii and cores were compiled from~\cite{Almeida:2024cqa},~\cite{2008ApJ...684.1075M,2018ApJ...860...66M} and~\cite{DES:2015zwj}, respectively. The last column shows the FDM mass estimated in this work.}
    \label{tab:table_1}
\end{table}

To convert the stellar mass-stellar core radius observations into constraints on the DM properties, we first assume that the DM core radius $r_c$ matches the stellar core radius $r^{*}_c$ with an uncertainty of $0.1\,{\rm dex}$, as suggested in~\cite{Almeida:2024cqa} using the Eddington inversion method and assuming an isotropic velocity profile. This assumption was shown to hold well for the small galaxies ($r_c \lesssim 0.5 \rm kpc$)~\cite{Teodori:2025rul} considered here. Following~\cite{Almeida:2024cqa}, we define the stellar core radius via $\Sigma(r^{*}_c) = \frac{1}{2} \Sigma(0)/2$, where $\Sigma$ denotes the projected surface density of the stellar profile. For the FDM halos, we use the results from Ref.~\cite{Chan:2021bja} where the DM profile is a broken power law that describes the core and transitions to the NFW profile. The core radius used in~\cite{Chan:2021bja} closely matches the core radius defined from half the projected surface density of DM as in~\cite{Almeida:2024cqa}.

\begin{figure}
  \centering
    \includegraphics[width=\columnwidth]{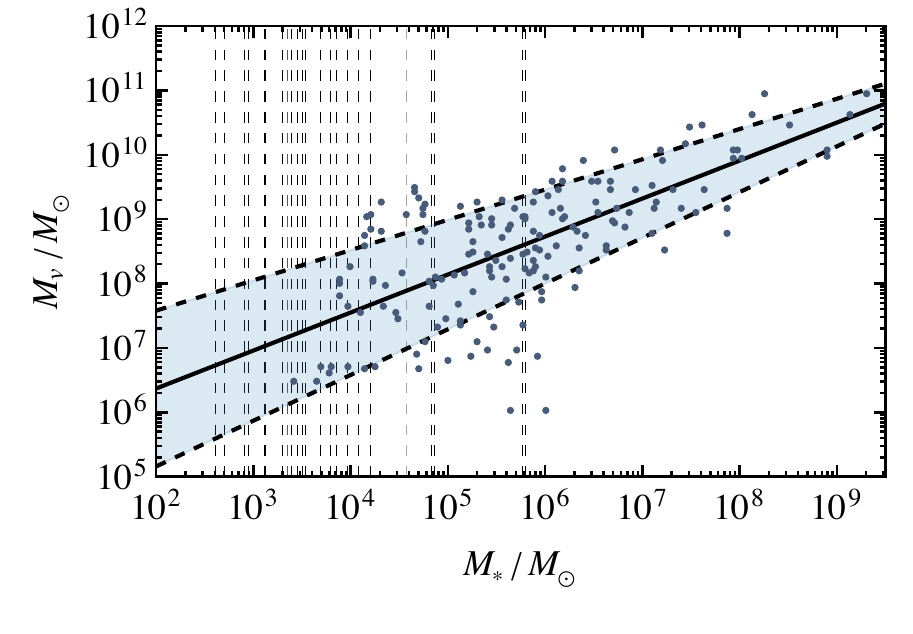}
    \vspace{-10mm}
    \caption{Stellar mass-halo mass relation for dwarf satellites compiled in~\cite{2021ApJ...923...35M,2024ApJ...961..236C} from hydrodynamical simulations~\cite{Bellovary:2018gbb,2021ApJ...923...35M,Applebaum_2021,2021ApJ...909..139A,2017MNRAS.470.1121T}, compared to the best fit using the fitting function~\eqref{eq:MvMsfit}. The shaded region shows the $1\sigma$ range of the distribution~\eqref{eq:MvMsfit} and the vertical dashed lines show the stellar masses of the UFDs listed in Table~\ref{tab:table_1}.}
    \label{fig:massrelation}
\end{figure}

Second, we relate the stellar mass to the DM halo mass using data of dwarf galaxies in zoom-in hydrodynamical simulations~\cite{Bellovary:2018gbb,2021ApJ...923...35M, Applebaum_2021,2021ApJ...909..139A,2017MNRAS.470.1121T}, as compiled in~\cite{2021ApJ...923...35M,2024ApJ...961..236C}. In particular, we consider satellite dwarf galaxies, which tend to have more stars for a given halo mass than the isolated ones~\cite{2024ApJ...961..236C}. These simulations assume CDM, but the stellar-to-DM halo mass ratio remains largely independent of the DM model, as different DM types primarily redistribute the available total halo mass~\cite{2021MNRAS.501.4610R,2025MNRAS.536.3338C}. We model the relation between the DM halo mass $M_v$ and the stellar mass $M_*$ by the probabilistic ansatz
\be \label{eq:MvMsfit}
    p(M_v|M_{*}) \propto M_v^{-1}\exp\bigg[- \frac{\log_{10}^2(M_v/\bar{M}_v(M_*))}{2 \sigma_{M_v}^2(M_*)} \bigg]\,,
\ee
where $\bar{M}_v(M_*)$ and $\sigma_{M_v}(M_*)$ are parameters characterizing the mean and the scatter of the $M_v-M_{*}$ relation.\footnote{There is a slight abuse of notation here. For a log-normal distribution, $\log_{10}\bar{M}_v/\msun$ and $\sigma_{\bar{M}_v}$ give the mean and variance of $\log_{10}M_v/\msun$.} We model them as power-laws,
\bea \label{eq:fit}
    \log_{10}\frac{\bar{M}_{v}}{\msun} &= \bar{M}_{v,0} +\bar{M}_{v,1}\log_{10} \!\frac{M_{*}}{10^9\,\msun} \,,
    \\
    \sigma_{M_v} &= \sigma_{M_v,0} + \sigma_{M_v,1}\log_{10} \!\frac{M_{*}}{10^9\,\msun} \,,
\eea
and find the best-fit parameters $\bar{M}_{v,0}=10.5$, $\bar{M}_{v,1}=0.59$, $\sigma_{M_v,0}=0.37$ and $\sigma_{M_v,1}=-0.12$. The simulated data points and the corresponding fit are shown in Fig.~\ref{fig:massrelation}. The grey-shaded area shows the $1\sigma$ region, which is seen to get wider towards lower stellar masses. Such an uncertain relation between the stellar and DM halo masses for light haloes will contribute significantly to the uncertainties in the implied FDM particle mass. Fig.~\ref{fig:massrelation} also demonstrates that simulations reach observed UFD masses considered here.

We relate the FDM particle mass $m_a$ to the DM core radius $r_c$ predicted as a function of the DM halo mass $M_v$. This relation is subject to significant scatter due to differences in merger histories and environments that a halo can have. We take the fits to FDM cosmological simulations,\footnote{The core radius-halo mass relation has also been estimated analytically~\cite{Schive:2014hza}. This analytical estimate agrees with numerical simulations~\cite{Taruya:2022zmt}, especially for lower masses $M_v < 10^9\, \msun$, which are the focus of our study.} carried out in Ref.~\cite{Chan:2021bja} which found that, at $z\approx 0$, the DM core radius is
\be \label{eq:Mcrc}
    \frac{r_c}{\rm kpc} = \left[\frac{m_a}{10^{-21}{\rm eV}}\right]^{-2}\left[ \frac{M_c}{5.5\times10^5 \msun}\right]^{-1}\,,
\ee
where $M_c$ denotes the DM core mass. To model the relation between the halo mass and core mass, we will adopt a probabilistic model instead of a deterministic one. The model is motivated by the result of Ref.~\cite{Chan:2021bja}, where a fitting function of the form
\bea\label{eq:McMv0}
    \frac{M_c(\vect{\theta})}{\msun} 
    &= 10^6 \,\beta \left[\frac{m_a}{8\times10^{-23}\,{\rm eV}}\right]^{-\frac32} \\
    & + 10^{-\alpha\gamma}\left[\frac{M_v}{\msun}\right]^{\alpha}\left[\frac{m_a}{8\times10^{-23}\,{\rm eV}}\right]^{\frac32 (\alpha-1)}\,,
\eea
was proposed. Above, $\vect{\theta}=(\alpha,\beta,\gamma)$ denotes the parameters. By fitting the simulations of~\cite{Nori:2020jzx, Schive:2014hza, Mocz:2017wlg}, Ref.~\cite{Chan:2021bja} found that upper and lower bounds of the scatter are given by $\vect{\theta_+}=(0.645\,, 8.52\, ,-3.35\, )$ and $\vect{\theta_-}=(0.326\,, 2.00\,,-14.11\,)$. However, as discussed in~\cite{Chan:2021bja}, the spread in the data is not a statistical uncertainty but rather a physical feature that arises because different haloes follow different relations due to their history and properties. To encapsulate this, we include the spread in the $M_c$-$M_v$ relation and model it by a log-normal distribution with average $\bar{M}_c(M_v,m_a) = \sqrt{M_c(\vect\theta_+) M_c(\vect\theta_-)}$ and standard deviation $\sigma_{M_c}(M_v,m_a) = \log_{10}[M_c(\vect\theta_+)/M_c(\vect\theta_-)]/4$,
\be \label{eq:McMv}
    p(M_c|M_v;m_a) \propto M_c^{-1}\exp\bigg[- \frac{\log_{10}^2(M_c/\bar{M}_c)}{2\sigma_{M_c}^2} \bigg]\,.
\ee

Finally, combining Eqs.~(\ref{eq:MvMsfit}, \ref{eq:Mcrc}, \ref{eq:McMv}), the DM core radii distribution for a given stellar mass and FDM particle mass is
\bea \label{eq:rcma_FDM}
    p_{\rm FDM}(r_{\rm c}|M_* ; m_a)
    &= \int \td M_v \,\td M_c \, \delta(r_c - r_c(M_c)) \\
    &\qquad \times p(M_c|M_v;m_a) p(\!M_v|M_*)\,.
\eea
Eq.~\eqref{eq:rcma_FDM} constitutes the generative probabilistic model for the $r_c-M_*$ relation in FDM models that will be confronted with the UFD data. 

The inverse proportionality of core size to core mass, as described above, is characteristic of the non-interacting FDM case. Although not explored in this study, relaxing this assumption and allowing for self-interactions enables a much broader range of qualitatively different behaviours~\cite{Chavanis:2011zi, Chavanis:2011zm, Glennon:2020dxs, Jain:2023tsr, Painter:2024rnc}. 
In particular, in the presence of significant repulsive self-interactions, the quantum core can attain a unique radius determined solely by the interaction strength and the DM particle mass, \emph{i.e.}, independent of $M_c$~\cite{Chavanis:2011zi}.

For model comparison, we will also consider a generic phenomenologically motivated generating model for the $r_{\rm c}-M_{*}$ relation given by the probabilistic power-law ansatz
\bea \label{eq:rcma_gen}
    p_{\rm ph}(r_{\rm c}|M_{*};\alpha,\beta,\sigma)
    &\propto
    r_c^{-1}\exp\bigg[- \frac{\log_{10}^2 (r_c/\bar r_{\rm c})}{2\sigma^2}\bigg] \, ,
\eea
where $\log_{10}(\bar r_{\rm c}(M_*;a,b)/{\rm kpc}) = \alpha + \beta \log_{10}(M_*/10^4 \msun)$ analogously to Eq.~\eqref{eq:fit}. In the FDM model, the scatter is practically independent of the stellar mass and thus we will omit this dependence also in the phenomenological model~\eqref{eq:rcma_gen}. In the presently relevant mass range $10^{-22}-10^{-19}$\,eV, the FDM model \eqref{eq:rcma_FDM} is very well approximated by the ansatz \eqref{eq:rcma_gen}, with the the parameters given by
\bea\label{eq:gen_FDM_matching}
    \mbox{FDM:} \quad
    \left\{\begin{array}{cc}
    \alpha &= -0.51 - 1.13 \mu_a - 0.03 \mu_a^2  \\
    \beta  &= -0.24 - 0.05 \mu_a - 0.01 \mu_a^2  \\
    \sigma &= +0.45 + 0.11 \mu_a - 0.02 \mu_a^2  
    \end{array}\right. \,,
\eea
where $\mu_a \equiv \log_{10}(m_a/10^{-21}\rm eV)$. Thus, the FDM scenario constitutes a one-parameter subset of the more general phenomenological generating model~\eqref{eq:rcma_gen}. 

\begin{figure*}[t]
    \centering
    \includegraphics[width=\textwidth]{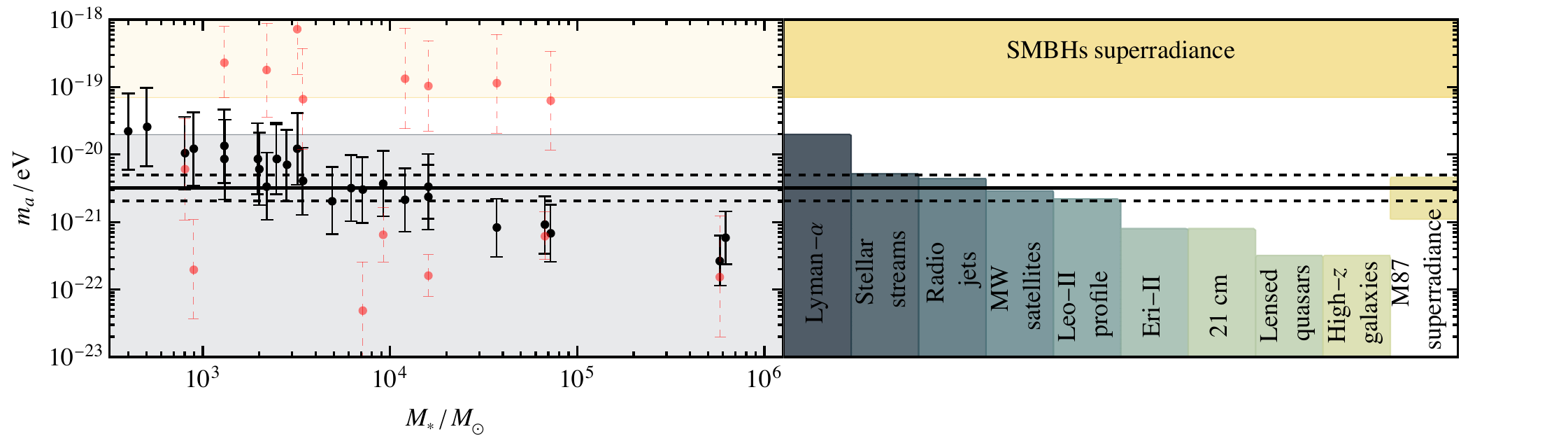}
    \vspace{-8mm}
    \caption{\textit{Left panel:} FDM particle mass fit for the UFDs listed in Table~\ref{tab:table_1}. The points show the best-fit values of $m_a$ and error bars in the $1\sigma$ confidence ranges. The red points with errors show the results presented in Ref.~\cite{Hayashi:2021xxu} with stellar masses taken from Table~\ref{tab:table_1}. \textit{Right panel:} Comparison of the fit of the UFD with astrophysical constraints on the FDM mass~\cite{Rogers:2020ltq, 2020PhRvD.101j3023B, Powell:2023jns,DES:2020fxi, Zimmermann:2024xvd,Marsh:2018zyw, Laroche:2022pjm,Kulkarni:2020pnb, Dalal:2022rmp}.}
    \label{fig:results}
\end{figure*}

\vspace{5pt}\noindent\textbf{Results -- } The predictions of the FDM model with stellar mass and DM core radii estimates are compared by the following likelihood for each UFD
\bea
    \mathcal{L}_j(\Lambda) &\propto 
    \int \td r_c\, \td M_*\,  p_M(r_c|M_*;\Lambda) p_j(r_c,M_*)\,,
\eea
where $p_M$ denotes the generating model parametrized by $\Lambda$. For FDM model~\eqref{eq:rcma_FDM} we have $\Lambda \in \{m_a\}$ and the general phenomenological model~\eqref{eq:rcma_gen} corresponds to $\Lambda \in \{\alpha, \beta, \sigma\}$. We adopt flat priors in the ranges $\alpha \in [-1.8,-0.6]$, $\beta \in [-1,1]$, $\sigma \in [0,1.5]$ and $\log_{10} m_a/{\rm eV} \in [-24,-16]$. The UFDs are labelled by  $j$ and $p_j(r_c, M_*)$ models the uncertainties in the stellar mass and the DM core radius, which we take to be uncorrelated and described by a log-normal distribution with the mean and variance adapted from Table~\ref{tab:table_1}. This is a conservative estimate as log-normal distributions tend to have longer tails than the normal distribution. Moreover, due to the probabilistic nature of the generating model, we find that the uncertainties in the inference of model parameters are dominated by the intrinsic spread in the generating model rather than uncertainties in the empirical estimates of stellar masses and core radii. Therefore, a more detailed account of the latter would have a marginal effect on the result. 

The preferred FDM particle masses for each of the UFDs are compiled in the right column of Table~\ref{tab:table_1} and in Fig.~\ref{fig:results} together with existing constraints on the FDM mass arising from SMBH superradiance~\cite{Stott:2018opm,Davoudiasl:2019nlo}, Lyman-$\alpha$ forest~\cite{Rogers:2020ltq}, stellar streams~\cite{2020PhRvD.101j3023B}, lensed radio jets~\cite{Powell:2023jns}, Milky Way satellites~\cite{DES:2020fxi}, density profiles of Leo~II~\cite{Zimmermann:2024xvd}, formation of Eridanus~II~\cite{Marsh:2018zyw}, 21\,cm observations~\cite{Nebrin:2018vqt}, lensed quasars~\cite{Laroche:2022pjm}, high-$z$ galaxies~\cite{Kulkarni:2020pnb} and FDM heating in Segue~I,II~\cite{Dalal:2022rmp} (not included in the plot). Combining all UFDs listed in Table~\ref{tab:table_1} gives $m_a = 3.2_{\scriptscriptstyle-0.6}^{\scriptscriptstyle+0.8} \times 10^{-21}\,$eV at $95\%$ CR. This is indicated by black horizontal lines in Fig.~\ref{fig:results}. 
These results are compatible with an earlier similar study based on 18 UFDs~\cite{Hayashi:2021xxu}.

\begin{figure}[b]
    \centering
    \includegraphics[width=\columnwidth]{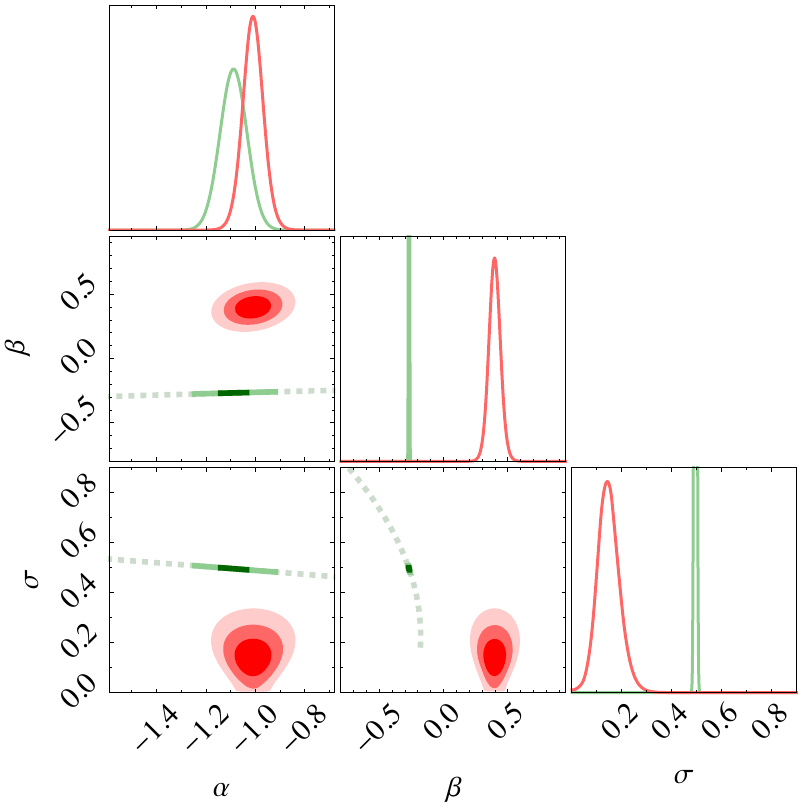}
    \vspace{-5mm}
    \caption{The red shading shows the $1,\,2,\,3\sigma$ CR regions of a power-law fit to the observed UFD stellar masses core radii shown in Table~\ref{tab:table_1}. The dashed green shows the parameter space of the FDM model \eqref{eq:gen_FDM_matching} in the range of the prior for $m_a$ and the solid green lines show the 1$\sigma$ and 3$\sigma$ CR for FDM.}
    \label{fig:const}
\end{figure}

Using Eq.~\eqref{eq:gen_FDM_matching}, the best fit axion mass corresponds to the following phenomenological $r_{\rm c}-M_{*}$ model~\eqref{eq:rcma_gen} parameters:
\bea
    \hspace*{-1pt}\mbox{FDM:} \,\,
    \alpha\!=\!-1.09{}^{\scriptscriptstyle +0.10}_{\scriptscriptstyle-0.11}, \,
    \beta \!=\!-0.27{}^{\scriptscriptstyle +0.01}_{\scriptscriptstyle -0.01}, \,
    \sigma\!=\!0.50{}^{\scriptscriptstyle +0.01}_{\scriptscriptstyle -0.01} \,,
\eea
The 1D and 2D marginalised posterior obtained for the phenomenological model is shown in Fig.~\ref{fig:const}. The best fit values for this model are
\be
    \hspace*{-7pt}\mbox{pheno:} \,
    \alpha\!=\!-1.01{}^{\scriptscriptstyle +0.04}_{\scriptscriptstyle -0.04}, \,\,\,\,
    \beta \!=\! 0.40{}^{\scriptscriptstyle +0.04}_{\scriptscriptstyle -0.04}, \,\,\,\,\,
    \sigma\!=\!0.14{}^{\scriptscriptstyle +0.04}_{\scriptscriptstyle -0.04} \,.
\ee
The $1\sigma$ and $3\sigma$ CR regions for the FDM scenario are depicted by the solid dark and light green curve in Fig.~\ref{fig:const}. Notably, the largest tension arises in the tilt of the $r_{\rm c}-M_{*}$ relation. However, we can also observe a sizeable disagreement in the variability $\sigma$ between the two models. We stress that, for FDM, the relation~\eqref{eq:gen_FDM_matching} essentially fixes $\beta$ and $\sigma$ in the relevant mass range so that their 1D posteriors in Fig.~\ref{fig:const} are essentially single lines. 

\begin{figure}
    \centering
    \includegraphics[width=\columnwidth]{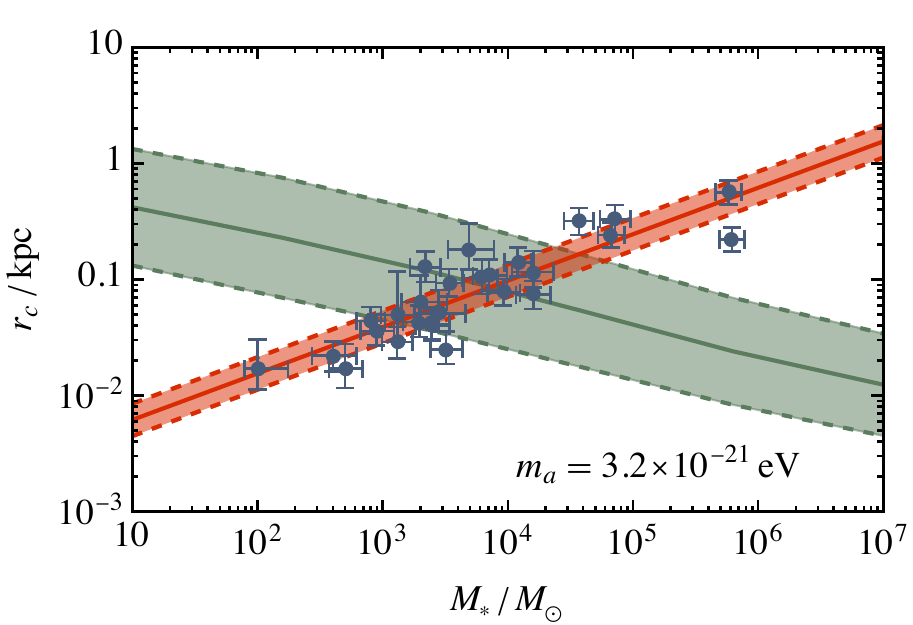}
    \vspace{-5mm}
    \caption{The best fits of the FDM model (green) and the phenomenological model (red), compared to the observations of dwarf galaxies (symbols with error bars) shown in Table~\ref{tab:table_1}. The coloured bands indicate the spread and contain 68\% of the predicted UFDs in the corresponding generating model.
    }
    \label{fig:galaxies}
\end{figure}

Fig.~\ref{fig:galaxies} shows the DM core mass-stellar mass relation of the best-fit FDM model and the best-fit phenomenological model alongside the observed UFDs. Together with Fig.~\ref{fig:const}, it illustrates clearly how observations prefer a positive correlation between stellar and core size ($\beta \gtrsim 0$), whereas FDM predicts a negative one ($\beta \approx -0.27$). This can also be observed in Fig.~\ref{fig:results} as heavier galaxies prefer a lighter FDM mass. The apparent conflict between the observed slope of the $r_{\rm c}-M_{*}$ relation and the prediction of FDM is well known and has been pointed out in the context of heavier galaxies~\cite{Hayashi:2021xxu, Burkert:2020laq, Safarzadeh:2019sre, Chan:2021bja}. However, this is the first time it has been quantified for UFDs.

Due to the matching \eqref{eq:rcma_gen}, the FDM model~\eqref{eq:rcma_FDM} can be interpreted as a nested hypothesis of the general phenomenological model~\eqref{eq:rcma_gen} allowing for a straightforward comparison. The log-likelihood difference between the two models is $\Delta\chi^2 \sim -2\log(\mathcal{L}_{\rm FDM}/\mathcal{L}_{\rm gen}) = 42.2$ indicating that the FDM hypothesis is excluded at $6.1\sigma$. Overall, we find that FDM is irreconcilable with UFD cores.

When considering individual UFDs, the smallest FDM mass is predicted by the heaviest UFDs in the sample, Canes Venatici-I requiring $m_a < 1.5\times10^{-21}\rm eV$ but also Draco, Hercules, Bo\"otes-I and Ursa Major-I for which $m_a \lesssim 5\times 10^{-21}\rm eV$ would be required. These UFDs imply a strong tension between an FDM origin of a DM core and the Lyman-$\alpha$ constraints as shown in Fig.~\ref{fig:results}. A tension between Canes Venatici-I and Lyman-$\alpha$ arising from a similar dynamical analysis was reported in Ref.~\cite{Hayashi:2021xxu}.

Since the FDM core would be largest in the lightest galaxies, the strongest constraints on the FDM mass must also arise from the lightest UFDs. Indeed, by Table~\ref{tab:table_1}, Cetus II imposes $m_a > 2.2 \times 10^{-21} \rm eV$ at the $2\sigma$ CL. Otherwise, its predicted FDM core size would exceed the observed one.

Finally, let us briefly consider the possibility that FDM is responsible for the core of a subset of the UFDs in Table~\ref{tab:table_1}, while the core in the remaining UFDs arises by some unspecified mechanism. The positive $r_c-M_*$ correlation found in the data and the negative correlation predicted by FDM then implies that FDM must be responsible for cored DM profiles in all UFDs below a given mass threshold. For example, assuming that only the 4 lightest UFDs with $M_* \lesssim 10^3 \msun$ have FDM cores would give $m_a = 2.2_{-1.1}^{+4.3} \times 10^{-20}\,$eV. Although this mass range lies at the edge of the viable window between superradiance and Lyman-$\alpha$ constraints (see Fig.~\ref{fig:results}), there is no good theoretical justification for omitting the heavier UFDs.

\vspace{5pt}\noindent\textbf{Conclusions -- } We studied the FDM hypothesis in the context of DM cores in 27 UFDs. The FDM scenario was modelled using a probabilistic generating function that was fitted to numerical simulation. We find that
\begin{itemize}[leftmargin=*]
    \item The preferred range of the FDM mass to explain our dataset lies in the range 
    \be
        2.6\times10^{-21}\,{\rm eV}<m_a<4\times10^{-21}\,{\rm eV}\,, 
    \ee
    which is in tension with existing bounds on $m_a$ arising, e.g. from Lyman-$\alpha$ observations, as illustrated in Fig.~\ref{fig:results}.
    
    \item While observations indicate a positive correlation between the DM core ratio to stellar mass, FDM predicts a negative one as illustrated in Fig.~\ref{fig:galaxies}. This extends the findings of~\cite{Burkert:2020laq, Salucci:2020cmt}, to much lighter stellar masses where it is believed that baryonic effects are not substantial. Compared to a generic phenomenological model for the DM core radius to stellar mass, the FDM scenario is excluded at a CL of $6.1\sigma$. 
\end{itemize}
We conclude that non-interacting FDM  cannot explain the core of these very light satellites, and if FDM cannot explain the smallest cores observed it fails to explain any other observed core since those would require even lighter DM masses. Independently of whether FDM is a viable explanation of DM cores in UFDs, these observations also constrain the axion mass and give a lower bound
\be
    m_a>2.2\times10^{-21}\,{\rm eV}
\ee
at the $2\sigma\, {\rm CL}$.

\vspace{2mm}
\begin{acknowledgements}
{\it Acknowledgements -- } We thank Kohei Hayashi for sharing results from his paper~\cite{Hayashi:2021xxu}. MB thanks Konstantin Karchev for his valuable insights and helpful discussions. This work was supported by the Estonian Research Council grants PSG869, PSG938, PRG1006, RVTT7, RVTT3, the Estonian Ministry of Education and Research (grant TK202) and the European Union's Horizon Europe research and innovation programme (EXCOSM, grant No. 101159513). KM was supported by the Estonian Research Council personal grant PUTJD1256. JSA research is partly funded by the Spanish Agencia Estatal de Investigaci\'on through grant PID2022-136598NB-C31(ESTALLIDOS 8) and by the European Union through the grant ''UNDARK'' of the widening participation and spreading excellence programme (project number 101159929). The work of VV was partially funded by the European Union's Horizon Europe research and innovation program under the Marie Sk\l{}odowska-Curie grant agreement No. 101065736.
\end{acknowledgements}

\bibliography{refs}

\end{document}